\documentclass[aps,pra,twocolumn,superscriptaddress,showpacs]{revtex4}

\usepackage{epsfig}

\begin{document}
\title{
Enhancing the capacity and performance of collective atomic quantum memory
}
\author{Tom\'{a}\v{s} Opatrn\'{y}}
\affiliation{Department of Theoretical Physics, Palack\'{y} University, 
17. listopadu 50, 77200 Olomouc, Czech Republic}

\author{Jarom\'{\i}r Fiur\'{a}\v{s}ek} 
\affiliation{Department of Optics, Palack\'{y} University, 17. listopadu 50,
77200 Olomouc, Czech Republic}

\date{\today}

\begin{abstract}
Present schemes involving the quantum non-demolition interaction between atomic
samples and off-resonant light pulses allow us to store quantum information
corresponding to a single harmonic oscillator (mode) in one  multiatomic
system.   We discuss the possibility to involve several coherences of each atom
so that the atomic sample can store information contained in several
quantum modes. This is achieved by the coupling of different magnetic sublevels of
the relevant hyperfine level by additional Raman pulses. 
This technique allows us to design not only the quantum
non-demolition coupling, but also beam splitterlike and two-mode squeezerlike
interactions between light and collective atomic spin.
\end{abstract}

\pacs{03.67.Mn, 
42.50.Ct, 
32.80.-t 
}

\maketitle


To enable quantum communication over long distances using quantum repeaters
\cite{Briegel98}, one needs to transform  a quantum signal from light pulses
into material media and vice versa.  Recently, this has been achieved by the
quantum non-demolition (QND) interaction between  alkali atom vapors and
off-resonant optical pulses \cite{JulsgaardNature01,FiurasekNature04} and also
by the light-induced transitions between atomic ground states  in a cloud of
atoms \cite{Wal03,Kuzmich03}. The principles of the off-resonant interaction
between light and polarized atoms were given in \cite{Happer67}, and recently
their application for collective spin measurements and quantum noise
suppression has been suggested \cite{Kuzmich98} and demonstrated
\cite{Kuzmich00a,Geremia04}.  It turns out that the QND scheme can be applied
to many valuable  quantum-information protocols \cite{Kuzmich00}.  However, for
most of them either the QND interaction has to be applied several times in
sequence (e.g., for swapping a quantum state between light and matter), or
measurements with feedback must be used (e.g., for transferring state from
light to matter that was  prepared in a well-defined initial state). This makes
many procedures rather cumbersome (one would have to store long optical
pulses to use the multi-pass schemes) so that more straightforward protocols
are highly desirable. Also, many questions remain as how to take full advantage
of the atomic degrees of freedom.

\begin{figure}
\centerline{\epsfig{file=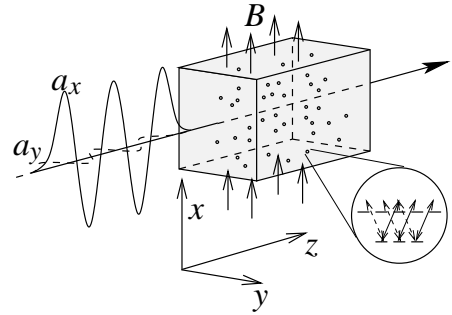,scale=0.5}}
\caption{\label{f-scheme1}
Geometry of the setup: light pulse travels in the $z$-direction, 
the $x$-polarization
mode is in a strong coherent state and the $y$-polarization is weakly
excited, carrying quantum signal. The pulse goes through atomic vapor placed in
magnetic field pointing in the $x$ direction, the light being detuned from
resonance of some electric dipole transition. 
}
\end{figure}

In this Letter, we propose a method to increase the amount of quantum
information  stored and processed in the atomic sample. We exploit the
structure of magnetic atomic levels and use  additional Raman coupling lasers
to store information  about several optical modes into various atomic
ground-state coherences. By properly choosing the frequency of the coupling
laser, it is possible to design various effective light-atom interaction
Hamiltonians such as QND coupling, two-mode squeezing  and beam splitter-like
coupling.  Especially the last one would be very useful to enable us quantum
state exchange between light pulses and atomic samples. The scheme could thus
greatly  enhance our ability  to process quantum information at the
light-matter interface.

\begin{figure}
\centerline{\epsfig{file=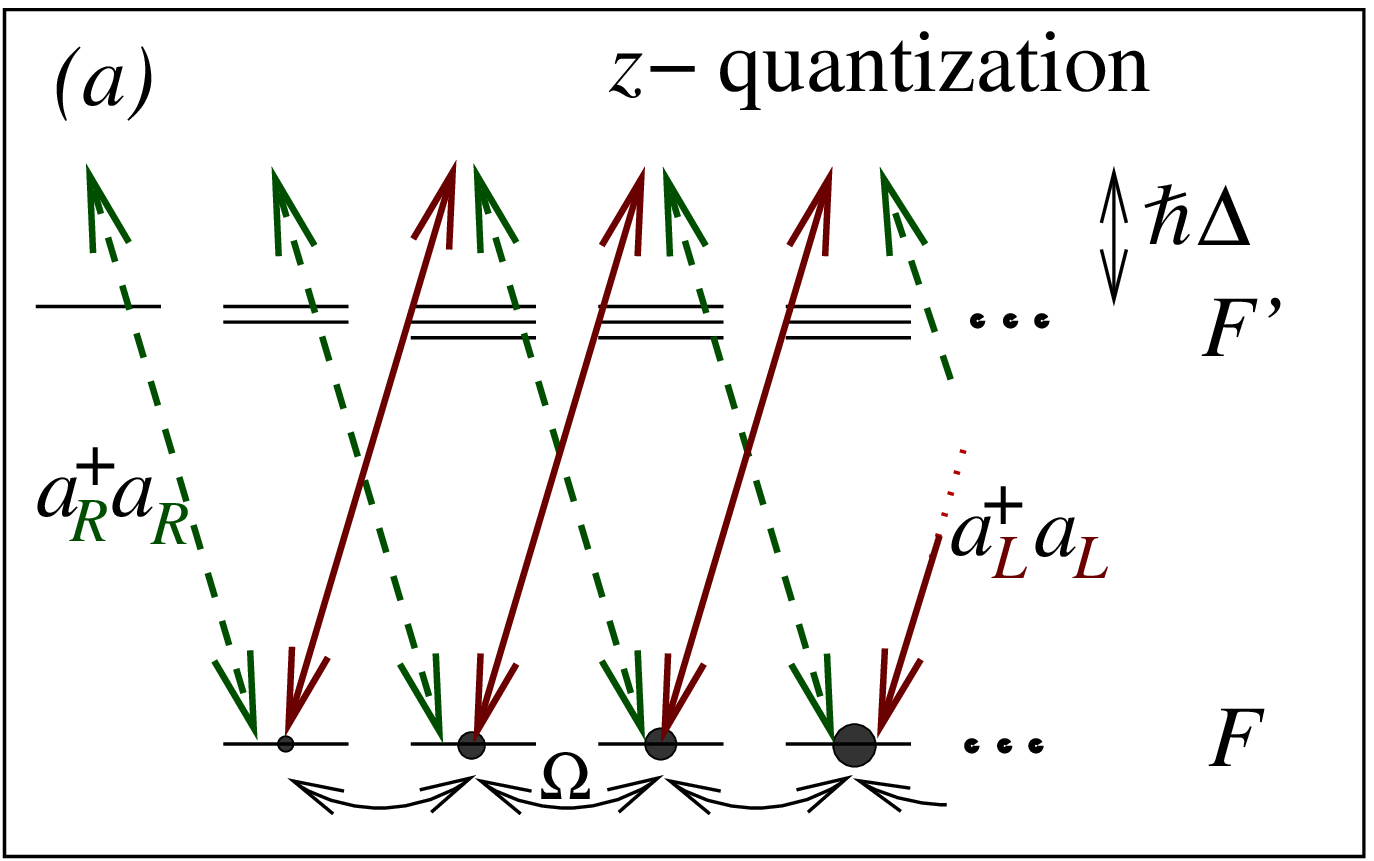,scale=0.40}}
\centerline{\epsfig{file=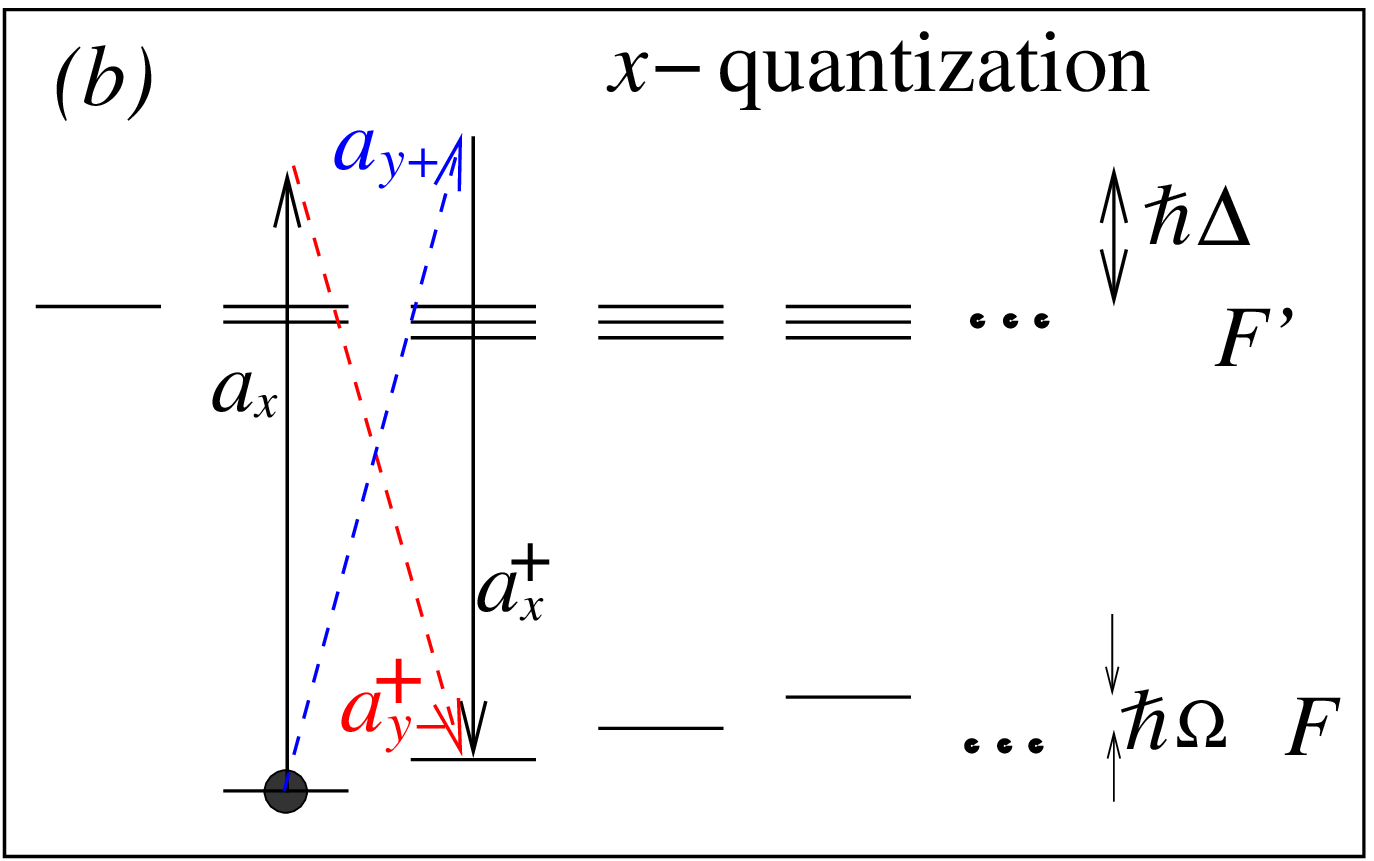,scale=0.40}}
\caption{\label{f-atlevels1}
Atomic level
scheme of the standard light-atom QND interaction using 
two different quantization axes. 
$(a)$: $z$-quantization, the arrows represent the 
$a_{R}^{\dag}a_{R}\sigma_{m,m}^{(z)}$
and
$a_{L}^{\dag}a_{L}\sigma_{m,m}^{(z)}$
terms of the Hamiltonian (\ref{ham2}).
$(b)$: $x$-quantization, the arrows represent
the $a_{x}a^{\dag}_{y-} \sigma_{-F+1,-F}$
and  $a_{y+}a^{\dag}_{x} \sigma_{-F+1,-F}^{(x)}$
terms of the Hamiltonian (\ref{hAFprime}). 
}
\end{figure}

The principle of the interaction is as follows. All atoms in the sample  are
initially pumped into a particular magnetic state $|m_F = -F\rangle_x$ of some
hyperfine level $F$ of the electronic ground state, with $x$ being the
quantization axis.  The information is then encoded into the coherence between
the state $|m_F = -F\rangle_x$ and  a weakly populated neighboring state  $|m_F
= -F+1\rangle_x$ by means of light  pulses traveling in the $z$ direction (see
Figs. \ref{f-scheme1} and \ref{f-atlevels1}b).  Each pulse has a strong
component which is $x$-polarized and a weak quantum component which is
$y$-polarized. With $z$ as the quantization axis, the influence of the field on
the atoms can be described as the ac Stark shift induced by the imbalance
between the right- and left-circularly polarized  components (see
Fig.\ref{f-atlevels1}a) changing the phases between different $|m_F\rangle_{z}$
states.  The influence of the atoms on the field can be understood as the
Faraday effect with asymmetry in the $\sigma_{m,m}^{(z)}$ populations causing
slight rotation of the linear polarization (here and in what follows
$\sigma_{m,n}$ denotes the operator $|m\rangle\langle n|$ involving the atomic
states $|m\rangle$ and $|n\rangle$ and the upper index denotes the quantization
axis).   With $x$ as the quantization axis, the interaction can be understood
as coherent Raman scattering  of the field on the  $\sigma_{-F,-F+1}^{(x)}$
coherences. In a typical experiment the atoms are placed in an $x$-oriented
magnetic field which causes rotation of the atomic polarization.  The effect of
the atoms on the optical field is then observed on the sideband with the Larmor
frequency $\Omega \propto B$, where $B$ is the magnetic induction.  In the
experiments so far, the information has been written into the  
$\sigma_{-F,-F+1}^{(x)}$ coherence  and its conjugate only, whereas all the
remaining coherences $\sigma_{-F,m}^{(x)}$ with $m\neq -F+1$ were negligible. A
natural question arises: could one use also these coherences as a medium for
storing quantum information?  The coherence $\sigma_{-F,m}^{(x)}$ would
contribute to the oscillation of  the atomic populations on the
$z$-quantization magnetic states with frequency $(m+F)\Omega$. Photodetection
on this sideband should reveal information about the corresponding coherence.

Let us first briefly discuss the traditional single-coherence QND scheme with a
multilevel atom interacting with an off-resonant field (see Fig.
\ref{f-atlevels1}).  The atom-field interaction Hamiltonian is $ H = \hbar
\sum_{m} \sigma^{(z)}_{m,m} (g_{mR}\mathcal{E}_R^{\dag}\mathcal{E}_R + g_{mL} 
\mathcal{E}_L^{\dag}\mathcal{E}_L) $, where $\sigma^{(z)}_{m,m}$ is the
population in the magnetic state  $|F,m\rangle_z$ with the quantization axis
$z$, and $\mathcal{E}_{R(L)}$  is the positive-frequency  operator of the
electric intensity of the  right (left) circularly polarized field. The
coupling constants are  $g_{mR(L)}=2/(\hbar^2
\Delta)\sum_{F'}\mu_{Fm,F'm\mp1}^2$ where $\Delta$ is the detuning from a given
fine-structure level (assumed to be much larger than the hyperfine splitting in
that level and much smaller than detuning from any other atomic level),
$\mu_{Fm,F'm'}$ is the dipole moment element between the hyperfine state
$|F,m\rangle$ of the ground electronic state and the hyperfine state
$|F',m'\rangle$ of the relevant excited electronic state, and the summation
runs over all hyperfine levels $F'$ in the fine-structure level.  In 
\cite{JulsgaardNature01,FiurasekNature04,JulsgaardThesis03,Julsgaard04}, 
cesium atoms are prepared in the hyperfine level $F=4$ of the atomic ground
state $6S_{1/2}$ and the field  is coupled to the transition to states on the
$6P_{3/2}$ level; the dipole moments sum up as $\sum_{F'}\mu_{Fm,F'm\pm
1}^2=\mu_0^2(8\pm m)/48$, where  $\mu_0^2=e^2|\langle
6S_{1/2}||r||6P_{3/2}\rangle|^2$  and the reduced matrix element can be
expressed in terms of the decay rate  $\gamma$ of the $6S \leftrightarrow 6P$
transition, $|\langle 6S_{1/2}||r||6P_{3/2}\rangle|^2=3c^2\gamma/(\alpha
\omega_0^3)$, where  $\alpha$ is the fine structure constant
\cite{JulsgaardThesis03}. Thus, the relevant part of the Hamiltonian is
\begin{eqnarray}
 H_{AF}=-\frac{\mu_0^2}{24 \hbar \Delta}
 (\mathcal{E}_{R}^{\dag}\mathcal{E}_{R} 
 - \mathcal{E}_{L}^{\dag}\mathcal{E}_{L})\sum_{m}m\sigma^{(z)}_{m,m} .
 \label{ham2}
\end{eqnarray}
We assume that the strong coherent $x$-polarized field has frequency
$\omega_0$. The atoms then resonantly couple to two $y$-polarized field modes
$a_{y\pm}$  oscillating on the Larmor sidebands $\omega_0\pm \Omega$ and we can
write $\mathcal{E}_{R,L}=E_0[a_x\pm i(a_{y+}+a_{y-})]/\sqrt{2}$, where $E_0$ is
the vacuum electric field, $E_0^2=\hbar \omega_0/(2 \epsilon_0 V)$,  $V=AcT$ is
the quantization volume, $A$ and $T$  are the transversal area and duration of
the optical pulse, respectively, and $a_x$ is the annihilation operator  of the
strong $x$-polarized field. It is convenient to work with nonmonochromatic
modes $a_{yC}=2^{-1/2} (a_{y-}+a_{y+})$ and $a_{yS}=2^{-1/2}(a_{y-}-a_{y+})$,
whose field quadratures $X_j = 2^{-1/2}(a_j + a^{\dag}_j)$, $P_j =
-i2^{-1/2}(a_j-a^{\dag}_j)$ are measured using homodyne detection as the sine
and cosine signal components oscillating at frequency $\Omega$. Transforming
from the $z$-quantization to the $x$-quantization, and averaging out fast
oscillating terms, the interaction Hamiltonian becomes

\begin{equation}
 \tilde H_{AF}^{(1)} = i\hbar
 \sum_{m}a_x^{\dag}G^{(1)}_m
 (a_{y+}\sigma_{m+1,m}^{(x)} + a_{y-}\sigma_{m,m+1}^{(x)}) + h.c.,
 \label{hAFprime}
\end{equation}
where $G^{(1)}_m=\mu_0^2 E_0^2/(48 \hbar^2 \Delta)\sqrt{20-m(m+1)}$. When the
$a_x$ mode is in a strong coherent state $|\alpha_0\rangle$ with $\alpha_0$
real and the atoms are initially prepared in state $|F,m=-F\rangle$ so that
only the coherences $\sigma^{(x)}_{-F,m}$ and  $\sigma^{(x)}_{m,-F}$ are
non-negligible,   after summing over all $N_A$ atoms, the interaction
Hamiltonian $H_{\rm int}^{(1)}  = \sum_{k=1}^{N_A} \tilde H_{AF,k}^{(1)}$ 
becomes
\begin{eqnarray}
 H_{\rm int}^{(1)}=\hbar \kappa ( P_C X_A + X_S P_A) .
 \label{QND}
\end{eqnarray}
This is the QND-type Hamiltonian in which
the atomic quadratures are
$ X_A = \frac{1}{\sqrt{2 N_A}} \sum_{k=1}^{N_A}(
\sigma^{(x,k)}_{-F,-F+1} + \sigma^{(x,k)}_{-F+1,-F})$, 
$ P_A = \frac{-i}{\sqrt{2 N_A}}\sum_{k=1}^{N_A}(
\sigma^{(x,k)}_{-F,-F+1} - \sigma^{(x,k)}_{-F+1,-F})$,
the coupling constant is $\kappa = -E_0^2\mu_0^2 \sqrt{N_L N_A}/(12\hbar^2 \Delta)$,
with the photon number $N_L=|\alpha_0|^2$, and the index $k$ denotes individual
atoms.

\begin{figure}
\centerline{\epsfig{file=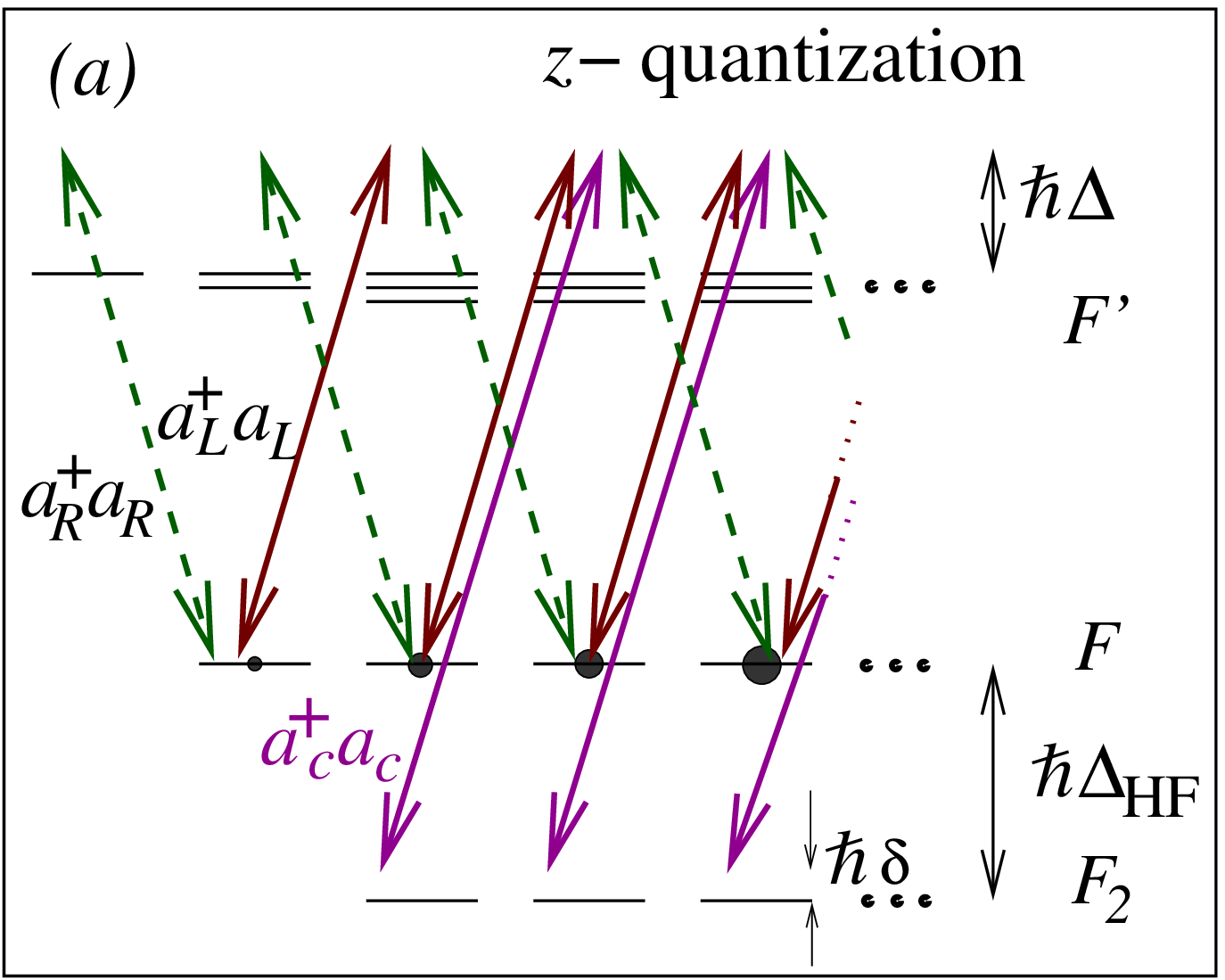,scale=0.405}}
\centerline{\epsfig{file=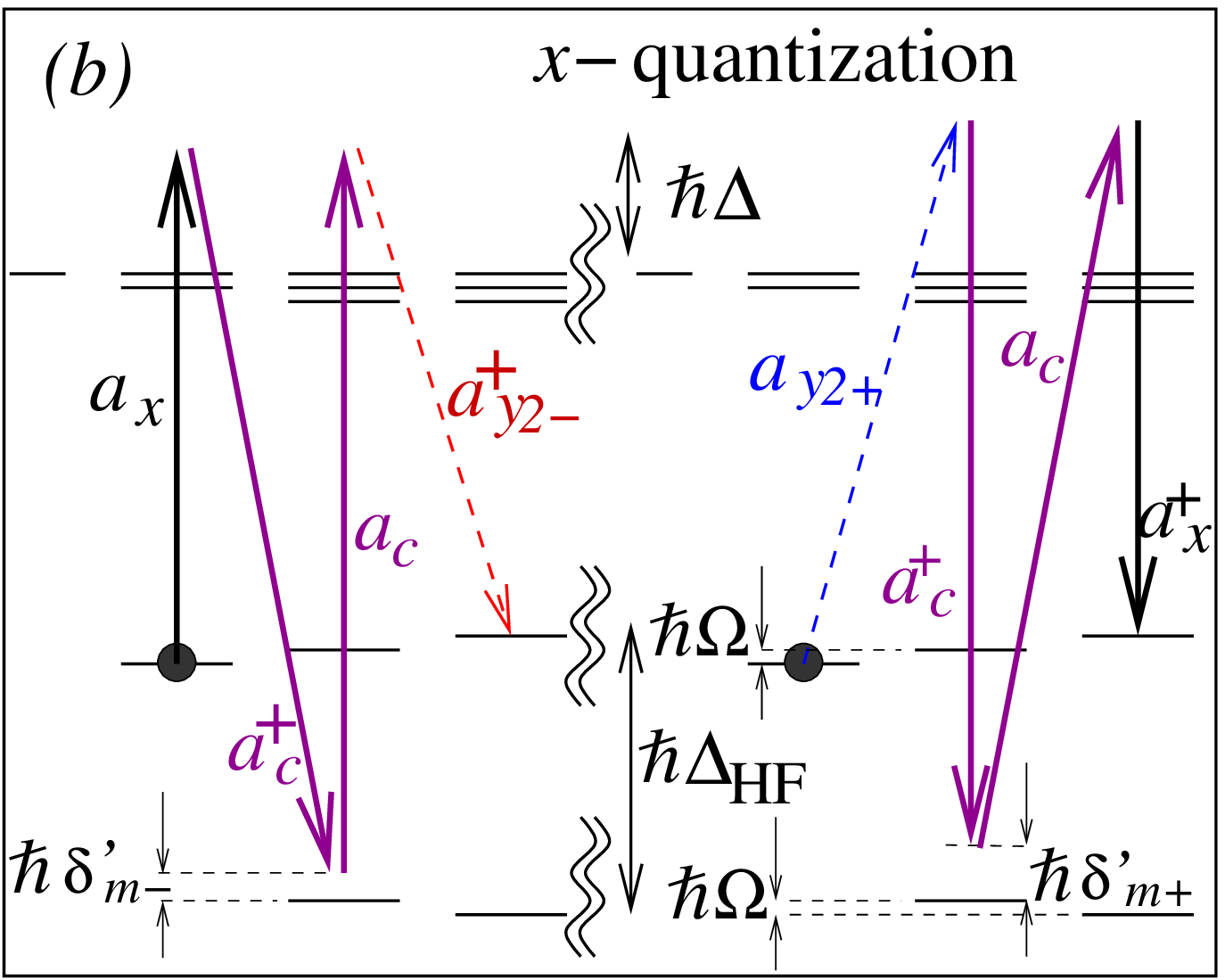,scale=0.405}}
\caption{\label{f-atlevels2}
Atomic level
scheme of the proposed light-atom  interaction for encoding the information
into the $\sigma_{-F,-F+2}$ coherence. $(a)$: 
$z$-quantization, 
$(b)$:
$x$-quantization, the arrows
represent the  $a_{x}a^{\dag}_c a_ca^{\dag}_{y2-} \sigma_{-F+2,-F}$ term 
and the $a^{\dag}_{y2-}a_c a^{\dag}_c a^{\dag}_{x}\sigma_{-F+2,-F}$ term
of the
Hamiltonian (\ref{HAF2x}). 
}
\end{figure}

To couple the light field to higher coherences we propose applying 
additional strong fields co-propagating with the signal fields.
A single left-circularly
polarized beam propagates in the $z$-direction, with frequency
$\omega_c$ slightly
off-resonance with respect to the Raman transition to the other hyperfine level 
$F_2$ of the ground state, $\omega_c-\omega_0 = \Delta_{\rm HF}-\delta$ 
(see Fig. \ref{f-atlevels2}). 
In the $z$-quantization picture,  
such a field enhances interaction between the atom and the
right-circularly polarized signal which is coupled to the
moment $\sum_{m}m^2\sigma_{m,m}^{(z)}$ oscillating at frequency $2\Omega$.
The Hamiltonian is most straightforwardly derived in the $x$-quantization.
If the  coupling field annihilation operator  is
$a_c$, we get after adiabatic elimination of the upper states and of the
lower $F_2=3$ state the additional interaction Hamiltonian in the form
\begin{eqnarray}
 H_{AF}^{(2)} = i\hbar
 \sum_{m}G^{(2)}_{m+}
 a_x^{\dag}
 a_c a^{\dag}_c a_{y2+}
 \sigma_{m+2,m}^{(x)} \nonumber \\ 
   + 
  i\hbar
 \sum_{m}G^{(2)}_{m-}
 a_x^{\dag}
 a_c a^{\dag}_c a_{y2-}
 \sigma_{m,m+2}^{(x)}
 + h.c.
 \label{HAF2x}
\end{eqnarray}
Here the $y$-polarized fields $a_{y2\pm}$ have frequencies 
$\omega_0 \pm 2\Omega$ and the couplings are
$ G^{(2)}_{m\pm} = E_0^2 E_{c0}^2 M^{(4)}_m/(\hbar^4 \delta'_{m\pm} \Delta^2)$,
where $M^{(4)}_m$ stands for the sum of all products of the dipole moment
elements relevant in the transition from 
$|F,m\rangle$ to 
$|F,m+2\rangle$ via the intermediate states in the
$P_{3/2}F'$, $S_{1/2}F_2$, 
and $P_{3/2}F'$ levels. 
For cesium with $F=4$ we find $M^{(4)}_m=-\mu_0^4
\sqrt{(3-m)(4-m)(5+m)(6+m)}/48^2$.
The Stark-shifted detunings $\delta'_{m\pm}$ are
\begin{eqnarray}
 \delta'_{m\pm} = \delta + \frac{\mu_0^2}{3\hbar^2 \Delta}
 (E_x^2 - E_c^2) + (2m+2\pm 1)\Omega ,
\end{eqnarray}
where $E_x$ and $E_c$ are the electric intensities of the $x$-polarized field
and of the coupling field. The second term
and the hermitian conjugate of the first 
term of the Hamiltonian (\ref{HAF2x}) 
with $m=-F$
are shown in Fig.~\ref{f-atlevels2}b.

If  $2\Omega \ll |\delta'_{m\pm}|$, 
then the $G^{(2)}_{m\pm}$ couplings are almost equal, 
$G^{(2)}_{m+} \approx G^{(2)}_{m-}$, and Eq. (\ref{HAF2x}) leads to a similar
expression as Eq. (\ref{QND}), now coupling field quadratures at the
$2\Omega$ sidebands to the atomic quadratures of the 
$\sigma_{m,m\pm 2}$ coherences. This would enable us to apply the QND
interaction similar to that in  
\cite{JulsgaardNature01,FiurasekNature04} in an additional channel.
Even more interesting results are obtained if the detunings $\delta'$
are smaller than $\Omega$ and the couplings 
$G^{(2)}_{m\pm}$ for  transitions $a_xa_c^{\dag}a_c a_{y2-}^{\dag}$ and 
$a_x^{\dag}a_ca_c^{\dag}a_{y2+}$ become very different.
Let us assume that the atoms were initially prepared in state
$|F,m=-F\rangle_{x}$ and let the coupling field be coherent with amplitude
$\alpha_c$ and the $a_x$ mode be coherent with real amplitude $\alpha_0$.   
If $|\delta'_{m=-F;-}| \ll |\delta'_{m=-F;+}|$, then
the dominant term in Eq. (\ref{HAF2x}) becomes
\begin{eqnarray}
 H_{AF}^{(2,{\rm SQ})} = i\hbar
 G^{(2)}_{{\rm SQ}}\alpha_0 |\alpha_c|^2
  \left( a_{y2-} \sigma _{2-} - 
  a_{y2-}^{\dag} \sigma _{2+}
 \right) ,
 \label{SQ}
\end{eqnarray}
where $G^{(2)}_{{\rm SQ}} = G^{(2)}_{m=-F;-}$ and 
the atomic operators are $\sigma _{2-} = \sigma^{(x)}_{-F,-F+2}$
and 
$\sigma _{2+} = \sigma _{2-}^{\dag}$. 
Operator (\ref{SQ})
acts as a two-mode squeeze operator, simultaneously creating or annihilating two
quanta, one in the field 
$a_{y2-}$
and one in the atomic
magnetic states. This would allow a direct preparation of 
entangled states between light and the atomic medium.
If, on the other hand
$|\delta'_{m=-F;-}| \gg |\delta'_{m=-F;+}|$, then
the dominant term in Eq. (\ref{HAF2x}) is
\begin{eqnarray}
 H_{AF}^{(2,{\rm BS})} = i\hbar
 G^{(2)}_{{\rm BS}}\alpha_0|\alpha_c|^2
  \left( a_{y2+} \sigma _{2+} - 
  a_{y2+}^{\dag} \sigma _{2-}
 \right) ,
 \label{BS}
\end{eqnarray} 
where $G^{(2)}_{{\rm BS}} = G^{(2)}_{m=-F;+}$. This Hamiltonian  corresponds to
a beam splitter which exchanges excitations between the field  $a_{y2+}$ and
the atomic magnetic states.  This would be suitable for swapping quantum
information  between the field and the medium.

Let us define the  annihilation operator of the effective atomic mode
associated with the coherence between magnetic levels $m=-F$ and $m=-F+2$, $
a_{A2}= \frac{1}{\sqrt{N_A}}\sum_{k}\sigma_{-F;-F+2}^{(x,k)}$. The total
Hamiltonian is obtained by summing the single-atom contributions over all $N_A$
atoms. For the squeezer-type Hamiltonian (\ref{SQ}) we obtain 
\begin{eqnarray}
 H_{\rm int}^{(2, {\rm SQ})}&=& i\hbar \kappa^{(2)} \left(
 a_{A2}^\dagger a_{y2-}^{\dagger}- a_{A2} a_{y2-}
 \right) ,
 \label{SQ2}
\end{eqnarray}
and the  beam splitter-like Hamiltonian is
\begin{eqnarray}
 H_{\rm int}^{(2, {\rm BS})}&=& i\hbar \kappa^{(2)} 
 \left( a_{A2}^\dagger a_{y2+}
 -a_{A2}a_{y2+}^\dagger  \right).
 \label{BS2}
\end{eqnarray}
The coupling constant can be expressed as 
$\kappa^{(2)}=E_0^2 E_{c0}^2 \mu_0^4 \sqrt{7 N_L N_A}N_c/(576 \, 
\hbar^4 \delta'
\Delta^2)$,  and $N_c$ stands for the number of photons in the coupling field.
Note that the Hamiltonians operating on the different sidebands  approximately
commute with each other (their commutators being $\sim N_{A}^{-1/2}$ times
smaller than squares of the Hamiltonians themselves). This means that  the
processes in the different channels can run independently of each other so,
e.g., an entangled state of two light modes can be stored  in the medium and
then be read out again.

The Hamiltonian $H_{\rm int}^{(2, {\rm BS})}$ enables us to straightforwardly
store the quantum state of light into the atomic ensemble and to retrieve it
later on. If  $\kappa^{(2)}T=\pi/2$, where $T$ is the effective interaction
time, then the quantum states of the light and atoms will be mutually
exchanged. Note that the storage or retrieval would require only a single
passage of  the light beam through the atoms. Our method also does not involve
any  measurement followed by feedback and can achieve high fidelity without
prior  squeezing of atoms or light, in contrast to the protocol of Ref.
\cite{FiurasekNature04}.

We can see that the four-photon coupling increases with the intensity of the
coupling field and with decreasing the detuning $\delta'$. It becomes comparable
to the two-photon coupling if
$ \Omega_c^2 \approx 48 \Delta \delta'/\sqrt{7} $,
where $\Omega_c =  E_{c0}\alpha_c\mu_0/\hbar$ is the Rabi
frequency of the coupling field. The 
magnitude of the detuning $\delta'$ is limited from
below by Doppler broadening of the hyperfine frequency $\Delta_{\rm HF}$,
$|\delta'| \gg \Delta_{\rm HF}v/c$, where $v$ is the 
RMS thermal speed of the atoms
and $c$ is speed of light. 
Note that the much stronger Doppler shift of the single-photon transitions 
$\omega_0 v/c \sim 10^9$s$^{-1}$ does not cause any problem here since the
signal and coupling fields propagate in the same direction and their Doppler
shifts subtract in the two-photon and four-photon transitions. This is the same
trick which was used for achieving ultraslow group velocity in hot gases
\cite{Kash99}.
For cesium atoms used in the experiments
\cite{JulsgaardNature01,FiurasekNature04} the RMS thermal speed is $v\approx
100$m/s and the hyperfine splitting is $\Delta _{\rm HF} = 9.1$GHz so that 
$\Delta_{\rm HF}v/c \approx 3$kHz. If we take $\delta \approx 30$kHz and the
single-photon detuning  as in \cite{JulsgaardNature01,FiurasekNature04},
$\Delta \approx 700$MHz, we find $\Omega _c \approx 10^7$s$^{-1}$. This value 
corresponds to the light intensity $\sim$mW/cm$^2$ which  is
of the same order of magnitude as  the intensity of the $x$-polarized field used
in the experiments. This suggests that the proposed method should work with 
present experimental setups.

In principle, by cascading multiphoton Raman transitions using 
additional intermediate
levels, one should be able to realize Hamiltonians coupling higher sidebands
with the corresponding atomic coherences. Another option 
could be going closer
to resonance with one of the upper hyperfine levels $F'$. 
Such a scheme would be between the far
off-resonant QND schemes and the resonant EIT schemes with the field coupled to 
multipole coherences \cite{Andrey}, or the resonant scheme
for selective addressing of higher polarization moments \cite{Yashchuk-PRL-03}.

In conclusion, we have proposed a scheme for involving higher coherences of the
atomic Zeeman sublevels by means of multiple Raman transitions to store quantum
information carried by light. The scheme enables us to work with a broader class
of Hamiltonians than those of the QND type and opens a way to involve higher
amount of modes to be stored in parallel in the atomic media.

Many stimulating discussions with R. Filip,
B. Julsgaard, A.B. Matsko, J.H. M\"{u}ller, D. O'Dell
and E.S. Polzik
are acknowledged.
This work was supported by the EU under project COVAQIAL (FP6-511004),
by GA\v{C}R (202/05/0486), and by M\v{S}MT (MSM6198959213).



\begin{thebibliography}{99}


\bibitem{Briegel98}
H.J. Briegel, W. D\"{u}r, J.I. Cirac, and P. Zoller,
Phys. Rev. Lett. \textbf{81}, 5932 (1998);
L.M. Duan, M.D. Lukin, J.I. Cirac, and P. Zoller,
Nature (London) \textbf{414}, 413 (2001). 

\bibitem{JulsgaardNature01}
B. Julsgaard et al., 
Nature {\bf 413,} 400 (2001).


\bibitem{FiurasekNature04}
B. Julsgaard et al.,
Nature {\bf 432,} 482 (2004).

\bibitem{Wal03}
C.H. van der Wal \emph{et al.}, 
Science \textbf{301}, 196 (2003).


\bibitem{Kuzmich03}
A Kuzmich \emph{et al.}, 
Nature (London) \textbf{423}, 731 (2003);
D.N. Matsukevich and A. Kuzmich, Science \textbf{306}, 663 (2004).



\bibitem{Happer67}
W. Happer and B.S. Mathur,
Phys. Rev. Lett. {\bf 18,} 577 (1967).

\bibitem{Kuzmich98}
A. Kuzmich et al., 
Europhys. Lett. {\bf 42,} 481 (1998);
%
A. Kuzmich et al., 
Phys. Rev. A {\bf 60,} 2346 (1999);
%
K. M{\o}lmer,
Eur. Phys. J. D {\bf 5,} 301 (1999);
%
Y. Takahashi et al., 
Phys. Rev. A {\bf 60,} 4974 (1999).

\bibitem{Kuzmich00a}
A. Kuzmich, L. Mandel, and N.P. Bigelow, Phys. Rev. Lett. \textbf{85}, 1594 (2000). 

\bibitem{Geremia04}
J.M. Geremia, J.K. Stockton, and H. Mabuchi, Science \textbf{304}, 270 (2004). 


\bibitem{Kuzmich00}
A. Kuzmich and E.S. Polzik,
Phys. Rev. Lett. {\bf 85,} 5639 (2000);
%
L.M. Duan et al., 
ibid {\bf 85,} 5643 (2000);
A. Di Lisi and K. M{\o}lmer, 
Phys. Rev. A {\bf 66,} 052303 (2002);
S. Massar and  E.S. Polzik,
Phys. Rev. Lett. {\bf 91,} 060401 (2003);
J. Fiur\'{a}\v{s}ek,  Phys. Rev. A \textbf{68}, 022304 (2003);
A. Kuzmich and T.A.B. Kennedy,
Phys. Rev. Lett.  {\bf 92,} 030407 (2004);
%
J. Fiur\'{a}\v{s}ek et al.,
ibid {\bf 93,} 180501 (2004);
%
K. Hammerer et al., 
Phys. Rev. A {\bf 70,} 044304 (2004).




\bibitem{JulsgaardThesis03}
B. Julsgaard, 
\textit{Entanglement and quantum interactions with macroscopic gas samples},
Ph.D. thesis, Aarhus University, Denmark  (2003).

\bibitem{Julsgaard04}
B. Julsgaard et al.,
J. Opt. B: Quantum Semiclas. Optics {\bf 6,} 5 (2004).



\bibitem{Kash99}
M.M. Kash et al., 
Phys. Rev. Lett. {\bf 82,} 5229 (1999).

\bibitem{Andrey}
M.S. Zubairy et al., 
Phys. Rev. A {\bf 65,} 043804 (2002);
A.B. Matsko et al., 
Opt. Lett. {\bf 28,} 96 (2003);
A.B. Matsko  et al., 
Phys. Rev. A {\bf 67,} 043805 (2003).

\bibitem{Yashchuk-PRL-03}
V.V. Yashchuk et al., 
Phys. Rev. Lett. {\bf 90,} 253001 (2003).



\end{thebibliography}
\end{document}